\documentclass[12pt,a4paper]{iopjournal}
\usepackage[utf8]{inputenc}

\usepackage{lmodern}
\usepackage{ragged2e}
\justifying
\DeclareUnicodeCharacter{2009}{\,}

\expandafter\let\csname equation*\endcsname\relax
\expandafter\let\csname endequation*\endcsname\relax
\usepackage{amsmath, amssymb}

\renewcommand{\articletype}[1]{}

\usepackage[
style=authoryear,
giveninits=true,
uniquename=init,
natbib=true,
url=false,
doi=false,
]{biblatex}
\renewcommand\cite{\parencite}

\usepackage{graphicx}
\usepackage{placeins}

\usepackage{xcolor}
\usepackage{soul}

\newcommand{\tabref}[1]{Table~\ref{#1}}
\newcommand{\figref}[1]{Figure~\ref{#1}}

\newsavebox{\mybox}
\newcommand{\equalbold}[1]{\sbox{\mybox}{#1}\makebox[\wd\mybox]{\textbf{#1}}}
\usepackage{hyperref}

\begin{document}

\articletype{Paper} 

\title{Quantification of dual-state 5-ALA-induced PpIX fluorescence: 
methodology and validation in tissue-mimicking phantoms}

\author{
Silvère Ségaud$^{1,\dagger}$, 
Charlie Budd$^{1,\dagger}$, 
Matthew Elliot$^{1,2, \dagger}$,
Graeme J. Stasiuk$^{3}$, 
Jonathan Shapey$^{1,2}$, 
Yijing Xie$^{1,*}$, 
and Tom Vercauteren$^{1}$
}

\affil{$^1$Research Department of Surgical \& Interventional Engineering, School of Biomedical Engineering \&  Imaging Sciences, King's College London, London SE1 7EH, UK}

\affil{$^{2}$Department of Neurosurgery, King's College London Hospital NHS Foundation Trust, London SE5 9RS, UK}

\affil{$^{3}$Research Department of Imaging Chemistry \& Biology, School of Biomedical Engineering \& Imaging Sciences, King's College London, London SE1 7EH, UK}

\affil{$^\dagger$These authors contributed equally to this work and share first authorship.}

\affil{$^*$Author to whom any correspondence should be addressed.}

\email{yijing.xie@kcl.ac.uk}


\begin{abstract}

Quantification of protoporphyrin IX (PpIX) fluorescence in human brain tumours has the potential to significantly improve patient outcomes in neuro-oncology, but represents a formidable imaging challenge. Protoporphyrin is a biological molecule which interacts with the tissue micro-environment to form two photochemical states in glioma. Each exhibits markedly different quantum efficiencies, with distinct but overlapping emission spectra that also overlap with tissue autofluorescence. Fluorescence emission is known to be distorted by the intrinsic optical properties of tissue, coupled with marked intra-tumoural heterogeneity as a hallmark of glioma tumours. Existing quantitative fluorescence systems are developed and validated using simplified phantoms that do not simultaneously mimic the complex interactions between fluorophores and tissue optical properties or micro-environment. Consequently, existing systems risk introducing systematic errors into PpIX quantification when used in tissue. 
In this work, we introduce a novel pipeline for quantification of PpIX in glioma, which robustly differentiates both emission states from background autofluorescence without reliance on a priori spectral information, and accounts for variations in their quantum efficiency. Unmixed PpIX emission forms are then corrected for wavelength-dependent optical distortions and weighted for accurate quantification. Significantly, this pipeline is developed and validated using novel tissue-mimicking phantoms replicating the optical properties of glioma tissues and photochemical variability of PpIX fluorescence in glioma. Our workflow achieves strong correlation with ground-truth PpIX concentrations ($R^2 = 0.918 \pm 0.002$), demonstrating its potential for robust, quantitative PpIX fluorescence imaging in clinical settings. 

\end{abstract}

\section{Introduction}
Fluorescence guided surgery (FGS) utilising 5-Aminolevulinic acid (5-ALA) derived Protoporphyrin IX (PpIX) was pioneered in glioma surgery, where it has been shown to double the rates of gross total resection (GTR) in high-grade glioma (HGG) ~\citep{Picart2024UseStudy,Stummer2006Fluorescence-guidedTrial}. These are the most common and fatal group of primary brain tumours, characterised by brain invasion and diffuse infiltrative margins~\citep{Sung2021GlobalCountries}. 
In current treatment paradigms, maximising tumour resection whilst preserving neurological function is the most significant modifiable prognostic factor in HGG~\citep{Chaichana2014EstablishingGlioblastoma} and achieving GTR increases mean survival in low grade glioma (LGG) from 60 to 90 months~\citep{Hervey-Jumper2014RoleGliomas}. The relative risk of death and tumour progression at 5 years post surgery for GTR vs subtotal resection in LGG is 0.39 and 0.50 respectively~\citep{Brown2019ManagementMeta-Analysis}.

FGS as currently practised relies on sufficient PpIX accumulation within tumours to allow direct intraoperative visualisation by the operating surgeon using a neurosurgical microscope augmented by band-pass filters. 
PpIX  is characterised by a peak fluorescence excitation at 405nm, with distinct photochemical states seen in vivo with primary emission peaks at 620~nm and 635~nm respectively and a broader secondary peak at 705nm~\citep{Montcel2013Two-peakedGlioblastomas}.
These two states are commonly referred to as PpIX$_{620}$ and PpIX$_{635}$, owing to their respective peak emission wavelengths. Both are visible intraoperatively as a pink fluorescence~\citep{Stummer1998InAcid} but are indistinguishable with the naked eye. 
Selective concentration of PpIX in glioma has been hypothesised to result from increased synthesis~\citep{Pustogarov2017HidingCells} reduced efflux~\citep{Mischkulnig2022HemeGliomas}, and accumulation by tumour associated macrophages~\citep{Nasir-Moin2024LocalizationMicroscopy,Liu2024Single-cellGlioblastoma,Lang2023MappingAnalysis}. In practice, the use of FGS is limited at HGG infiltrative margins where tumour cell density is lower and recurrence is common~\citep{Petrecca2013FailureGlioblastoma}, with negative predictive values as low as 35\%~\citep{Schupper2021Fluorescence-GuidedSurgery}. In LGG only around 20\% of tumours show visible fluorescence~\citep{Jaber2016TheFactors,Jaber2019IsGliomas}. Such limitations are thought to contribute to the relatively modest improvements in patient survival obtained in current studies of PpIX FGS ~\citep{Stummer2006Fluorescence-guidedTrial, Picart2024UseStudy, Fountain2021IntraoperativeMeta-analysis}.

Accurate, sensitive and specific quantification of PpIX fluorescence has the potential to address this significant limitation, and is increasingly being explored for 5-ALA derived FGS \citep{Gautheron20245-ALAReview}.  
Regardless of the implementation, quantification of PpIX is achieved by a combination of two processing steps applied to the detected fluorescence signal. The first step involves correction of tissue-induced optical distortions to the detected signal, while the second step consists of the extraction of PpIX-specific signal from overlapping autofluorescence.~\citep{Kotwal2024HyperspectralApplications,Valdes2019QuantitativeNeurosurgery}.

In the context of PpIX quantification, optical distortion refers to the interaction of tissue intrinsic properties, i.e., absorption and scattering, and light at the excitation and emission wavelengths of PpIX. 
Optical distortion introduces bias in the estimation of PpIX accumulation in tissue when relying on conventional fluorescence imaging. Existing methods for correction of optical distortions utilise either empirical reflectance-based or model-based approaches~\citep{Valdes2019QuantitativeNeurosurgery} and are discussed in the next Section~\ref{Related work}.

The extraction of PpIX specific signal relies on unmixing PpIX from overlapping tissue autofluorescence, requiring a priori knowledge of the fluorophores contributing to the detected fluorescence signal. This implies an understanding of autofluorescence emission, and the complex photochemical behaviour of PpIX in glioma tissue where two distinct porphyrin emission forms with emission maxima at 620nm and 635nm have been demonstrated~\citep{Alston2019SpectralGliomas}. 
These have been hypothesised to represent either varying aggregations of PpIX~\citep{Montcel2013Two-peakedGlioblastomas,Alston2019SpectralGliomas,Black2021CharacterizationSurgery}, or emission from other porphyrin intermediaries or photobreakdown products in heme biosynthesis such as Coproporphyrinogen III~\citep{SueroMolina2023UnravelingBiology}. Significantly, the quantum efficiency of these emission states is markedly varied \citep{Alston2018NonlinearPhantoms}, introducing a risk of systemic bias into PpIX quantitation in glioma when performed without dual-state unmixing.

The development and validation of fluorescence quantification strategies requires the use of phantoms which replicate both the optical properties and spectral behaviour of PpIX in a glioma micro-environment simultaneously.
This has been challenging with PpIX where phantoms either do not replicate the two photo states seen in vivo, and are limited to one emission form~\citep{Walke2023ChallengesMeasurements}, or do not capture the range of optical properties encountered in glioma~\citep{Alston2018NonlinearPhantoms,Jacques2013OpticalReview,Shapey2022OpticalWindow}. 
Additionally, phantom-based calibration of the quantification pipeline is usually needed for quantification in absolute rather than relative terms, i.e. for providing measurements of PpIX accumulation in concentration metric units.
Limitations in existing phantoms is thus also problematic for calibration purposes, where changes in PpIX quantum efficiency and spectral behaviour between non biological phantoms and in vivo can significantly impact the extracted PpIX concentration~\citep{Lu2020FluorescencePhantoms,Walke2023ChallengesMeasurements}. 
The methodology and validation of existing strategies for PpIX quantification is discussed in Section \ref{Related work}.

This work describes the derivation and validation of a novel method of PpIX quantification, designed explicitly for translation into the clinical use case of glioma surgery. Critically the dual emission forms of PpIX known to exist in vivo are robustly unmixed without relying on a priori knowledge of their spectral signature, whilst simultaneously correcting for the known range of optical properties for both high and low grade glioma. Our proposed quantification method is both derived and validated using novel, well characterised, tissue-mimicking phantoms that replicate the photochemical fluorescent dual-emission and optical properties seen in glioma tissue.

\section{Related work}\label{Related work}

Accurate quantification of PpIX fluorescence for use in FGS has been under investigation since the technique was introduced into clinical practice following a landmark stage 3 clinical trial in 2006 \citep{Stummer2006Fluorescence-guidedTrial}.This has been attempted using both probe-based and wide-field spectroscopic techniques \citep{Gautheron20245-ALAReview}.  As outlined above, despite variations in imaging techniques and hardware, fluorescence quantification fundamentally requires the correction of tissue induced optical distortion and unmixing of overlapping fluorescence signals to isolate the contribution from PpIX. The resulting \emph{corrected} fluorescence is then calibrated against  a series of phantom with known PpIX concentrations to extract a measurements in concentration units. 

\subsection{Optical correction}
\label{sec:optcorrbkg}

Light transport in biological tissue is governed by the tissue intrinsic optical properties and geometry \citep{Jacques2013OpticalReview}. The propagation of light within tissue is mainly affected by two phenomena: light absorption, naturally induced by the presence of endogenous chromophores and light scattering, caused by the tissue microstructure. Both of these phenomena are wavelength-dependent, and are determined by tissue composition. These translate into tissue intrinsic optical properties, mostly described using the absorption coefficient $\mu_{a}$ and reduced scattering coefficient $\mu_{s}'$. 
The PpIX fluorescence process occurring within tissue is therefore affected by tissue-light interactions, over both excitation and emission wavelength ranges. Specifically, excitation light penetrates within tissue and undergoes absorption and scattering prior to reaching PpIX. Subsequently, the emitted fluorescence photons migrate from within the tissue to a detector. In a linear regime, the intensity of fluorescence emission is expected to be proportional to the fluorophore concentration \citep{Lu2020FluorescencePhantoms}. Distortions induced by tissue optical properties therefore affect the perception of fluorophore accumulation in tissue and subsequent viability assessment. Specifically, the same amount of fluorophore will be perceived as lower intensity in a higher absorbing and lower scattering medium, but higher intensity in a lower absorbing and higher scattering medium. 

Correction methods can be found in the literature and split in two main categories: Model-based corrections and empirical corrections \citep{Valdes2019QuantitativeNeurosurgery}. Model-based approaches use light transport models in tissue with strong assumptions on tissue composition and geometry to describe the impact of tissue on fluorescence excitation and detection of emission. These yield algorithms to compensate for optical distortions to be applied to raw fluorescence spectra \citep{Kim2010QuantificationMeasurements, Saager2011QuantitativeDomain, Sibai2015QuantitativeResection, Valdes2017QF-SSOP:Imaging}. Alternatively, empirical approaches use surrogates for tissue optical properties such as diffuse reflectance. These yield compensation of distortions in a simpler form of a tuned correction factor \citep{Valdes2012ASurgery, Xie2017Wide-fieldResection}. The derivation of this correction factor is commonly obtained by calibration against tissue-mimicking phantoms. 

Proposed implementations can be divided into two types: spectroscopic point-probes and widefield devices. Point-probe devices are based on a handheld bundle of fibres \citep{Kim2010QuantificationMeasurements, Valdes2012ASurgery, Bravo2017HyperspectralTumors}. These are coupled to light engines for either fluorescence excitation or broadband illumination of tissue. 
Dedicated fibres are coupled to spectrally-resolved detectors, or spectrometers, for detection of fluorescence emission and diffuse reflectance. Quantitative optical properties can be extracted from diffuse reflectance measured at multiple source-detector separations and used as input to distortion correction models. High sensitivities with PpIX concentrations down to $\sim$1 ng/mL are reported for such system
\citep{Valdes2019QuantitativeNeurosurgery}. 
However, the benefit of point systems is limited in clinical settings as they do not allow for real-time inspection of the entire surgical field though PpIX concentration mapping.
Widefield devices use spectrally-resolved imaging systems typically composed of a variable optical filter such as a liquid-crystal tunable filter (LCTF) and a monochrome camera \citep{Valdes2012QuantitativeImaging, Xie2017Wide-fieldResection}. The extraction of quantitative tissue optical properties has been demonstrated using spatial frequency domain imaging and used as input to achieve fluorophore quantification, at the cost of increased instrumentation complexity \citep{Saager2011QuantitativeDomain, Sibai2015QuantitativeResection, Sibai2017PreclinicalResection, Valdes2017QF-SSOP:Imaging}.
This approach is therefore more suited for empirical correction models using diffuse reflectance, which have been demonstrated in clinical settings. Despite the development of novel hyperspectral imaging devices, the spectral resolution of such systems is significantly lower than those of point-probe systems due to the trade-off the be operated between spatial, spectral and temporal resolutions. Reported PpIX detection thresholds are higher than point-probes with typical concentrations of $\sim$10-100~ng/mL~\citep{Valdes2019QuantitativeNeurosurgery}, when using highly sensitive cooled scientific complementary metal-oxide semiconductor (sCMOS) or electron-multiplying charge-coupled device (EMCCD) cameras. 

\subsection{Spectral unmixing}

Initial techniques for unmixing PpIX emission from measured fluorescence emission utilised linear regression with a priori assumptions for the number of constituents and their respective fluorescence emission spectra, called basis spectra. Measured spectra were generally assumed to represent linear combinations of known end-members. In the case of glioma tissue, these may include fixed emission profiles from PpIX, PpIX photoproducts and variably defined contributions from endogenous fluorophores. The latter may be grouped as composite constituents and referred to as autofluorescence, or background signal - in both point-probe based  \citep{Valdes2011Delta-aminolevulinicMalignancy, Montcel2013Two-peakedGlioblastomas} and widefield systems \citep{Bravo2017HyperspectralTumors, Kaneko2019Fluorescence-BasedGliomas, Black2024DeepSurgery}. 

As emerging spectroscopic studies increasingly revealed micro-environment induced dual-state PpIX emission in phantoms \citep{Alston2018NonlinearPhantoms} and glioma tissue \citep{Montcel2013Two-peakedGlioblastomas, Black2021CharacterizationSurgery}, unmixing models were refined to include PpIX$_{620}$ and PpIX$_{635}$ spectra along with autofluorescence. PpIX contribution to measured fluorescence could then be expressed as a ratio of each emission form  \citep{Alston2019SpectralGliomas}, and distinct contributions \citep{Black2021CharacterizationSurgery}. 
It is important to note that, in these studies, quantification of dual-state PpIX in concentration units was not possible as only one PpIX emission state was present in calibration phantoms. Hence each emission state was modelled as a relative contribution to the measured signal.

Most recently, deep learning techniques have been employed to refine the unmixing process, allowing a unified optical correction and unmixing pipeline to handle the non-linearity inherent in such a biochemically and optically complex system \citep{Black2024DeepSurgery}. 
Supervised and semi-supervised systems were developed using stacked white light and fluorescence emission spectra, and fixed end member spectra as inputs, showing good performance in single emission state phantoms and pig brain homogenates. 

Techniques that rely on pre-set basis spectra risk mischaracterising fluorophores, particularly in the use case of glioma surgery where both intra- and inter-tumoural heterogeneity are core characteristics of the pathology. 
Furthermore,  unmixing techniques that rely on predefined spectra also do not account for shifts in emission spectra known to be induced by the glioma microenvironment \citep{Cannon2021CharacterizationDehydrogenase} and cannot be be trained in phantoms where the constituents do not precisely mimic those seen in glioma. The constituents of autofluorescence and various photochemical forms of PpIX or its breakdown products present in glioma are currently under investigation~\citep{Schaefer2019NADHResearch,SueroMolina2023UnravelingBiology}.

\subsection{In vitro PpIX phantoms}
Tissue-mimicking phantoms with PpIX are widely used for validation and calibration of fluorescence quantification algorithms and instruments. Most phantoms found in the literature are liquid and include intralipids as scattering agent. Absorbing agents may vary and include yellow dyes \citep{Kim2010QuantificationMeasurements, Valdes2012ASurgery, Sibai2015QuantitativeResection, Reinert2019QuantitativeVisualization}, India ink \citep{Saager2011QuantitativeDomain}, red dyes \citep{Black2024DeepSurgery}, blood \citep{Bravo2017HyperspectralTumors} and haemoglobin \citep{Xie2017Wide-fieldResection}. 
The liquid form of phantoms allows for easy fabrication and multiplication of the number of phantoms by successively increasing concentrations of optical agents and PpIX.
Even though the mentioned absorbing agents can be employed to tune the phantoms optical properties at the excitation (405 nm) and emission (600-750 nm) wavelengths of PpIX, they do not provide a high-fidelity representation of tissue optical properties throughout the 400-750 nm wavelength range. 
As an alternative, solubilized tissues or blended tissues have been used and referred to as homogenates~\citep{Kim2010QuantificationMeasurements, Black2024DeepSurgery}. 
These homogenates are spiked with known concentrations of PpIX and used as phantoms. While they allow for high-fidelity reproduction of tissue absorption, covering a wide range of optical properties using homogenates is more challenging. Furthermore, the reproduction of tissue scattering is not guaranteed in homogenates as this process alters the tissue microstructure.

All these previous phantom approaches consider PpIX in a single 635 nm photo-state and focus on PpIX concentration ranges showing linear increase in signal with concentration.  
Replicating in phantoms the spectral complexity of PpIX fluorescence as observed in vivo has only been investigated in a small number of studies \citep{Montcel2013Two-peakedGlioblastomas, Alston2018NonlinearPhantoms, SueroMolina2023UnravelingBiology}. 

\section{Materials \& Methods}

\subsection{Phantom fabrication} 
The composition and manufacturing of the phantoms presented in this work were adapted from previously published work~\citep{Bahl2024AExperiments}. Phantoms were prepared with a gelatin concentration of 6\% by mass to yield firm phantoms for ease of handling.
Gelatin was dissolved in phosphate-buffered saline (PBS, Sigma Aldrich, UK) and heated with a magnetic stirrer hotplate to reach 50°C. 20\% stock intralipids (IL, Fresenius Kabi, Bad Homburg, Germany) and  
powdered haemoglobin (Hb, Haemoglobin from bovine blood, lyophilized powder, H2500-5G, CAS 9008-02-0, Sigma Aldrich, UK) were chosen as optical agents to mimic light scattering and absorption observed in biological tissue, respectively. Stock solutions of IL and Hb in 1x PBS were mixed thoroughly and paraformaldehyde (PFA) was added to a final concentration of 0.2\% by volume for stability. Once the gelatin dissolution was complete, the solution was cooled to 35°C and added to the solution with optical agents to reach the desired concentrations of IL and Hb.

A stock solution of 500 $\mu$g/mL PpIX (Protoporphyrin IX disodium salt, 258385, Sigma Aldrich, UK) in dimethyl sulfoxide (DMSO, Cambridge Bioscience, UK) was prepared and added last to the preparation. A silicone mould was placed on a bed of ice to allow quick setting of the gelatin. The final preparation was ultimately mixed, then poured into the mould and covered with aluminium foil to shield the phantoms from ambient light. Once the gelatin was set, phantoms were stored at 5°C in a fridge prior to mounting in custom holders. 
For this study, phantoms were prepared with the following concentration ranges: 1–5 \% IL, 0.5–2 mg/mL Hb, 0–15 $\mu$g/mL PpIX.  The PpIX concentration range was chosen to be close to clinical concentrations based on our previous study on freshly excised glioma specimen \citep{Elliot2024ANSURGERY}. 

Phantoms were mounted in custom holders prior to undertaking the measurements. A clear acrylic sheet of 1.5 mm thickness was cut into rectangular pieces featuring a 8 x 18 mm rectangular window. Phantom slabs were cut to these dimensions, mounted in this window and sandwiched between two glass coverslips. Coverslips were glued to the acrylic piece using an optical adhesive (Pertex, Histolab, Askim, Sweden). Once sealed, phantoms were labelled and stored at 5°C in the fridge prior to measurements.  
This produced mono-layered fluorescent tissue-mimicking phantoms with constant dimensions and homogenous distribution of optical agents and fluorophores.

\subsection{Fluorescence measurement}
Data collection for the characterisation of fluorescence properties was performed with a high-sensitivity spectrofluorometer (FS5, Edinburgh Instruments, UK), featuring a front-facing sample holder module (SC-10) compatible with mounted samples as previously described. Fluorescence emission spectra were measured under excitation from a filtered Xenon light source, centred at 405 nm with a fixed bandwidth of 1.5 nm. 
The sample position was adjusted to align the excitation beam with the centre of the phantom front-facing surface. 
Fluorescence emission was scanned by a built-in monochromation stage and photomultiplier tube (PMT) setup. An additional fluorescence emission long-pass filter (455FCS, Knight Optical) was added in the detection path within the sample holder module for increased rejection of excitation light. Emission spectra were collected over a 450–800 nm range with a 1 nm step, with fixed slit width of 1.5 nm and dwell time of 0.2 s. Built-in corrections for dark noise and excitation source fluctuations were enabled for all measurements.
In addition to built-in corrections, pre-processing of fluorescence emission scans included signal normalization for exposure (dwell time, scan repeats) and compensation for the additional emission filter transmission at scanning wavelengths. 
To mitigate the variation in measurements from the potential non-homogeneity, the measurements were repeated three times for each sample while changing the phantom position within the spectrofluorometer holder.

\subsection{Optical properties extraction}
Optical properties were measured as transmittance and reflectance using a high-performance dual-beam spectrophotometer (Lambda750s, Perkin Elmer, USA) featuring a single 100 mm integrating sphere with a pair of PMT and InGaAs detectors. This method and processing workflow used to extract optical properties from reflectance and transmittance were previously described for the optical characterisation of human brain and tumour tissue~\citep{Shapey2022OpticalWindow}. 
Total reflectance scans were acquired using an 8° wedged reflectance port cover and assorted custom sample holder. The instrument was calibrated prior to total reflectance measurements using a standard Spectralon target. Next, total transmittance scans were acquired using a custom sample holder over the transmittance port, with a standard Spectralon target covering the reflectance port. 
Both scans were performed over a 400–1500 nm wavelength range with a 10 nm step.
The primary beam size was reshaped for both reflectance and transmittance scans using dedicated combinations of optical components. An adjustable slit and a lens assembly were used to obtain a low-divergence beam focused on the surface of the sample, illuminating an area smaller than the phantom area and smaller than the ports. The beam size at the reflectance and transmittance ports was 2$\times$3.5 mm and 1$\times$1.5 mm, respectively.  

The extraction of optical properties was achieved using a 2-stage inverse adding-doubling (IAD) method described previously~\citep{Shapey2022OpticalWindow}. IAD is a well-suited inverse problem solver to determine intrinsic optical properties, namely the absorption coefficient $\mu_{a}$ and reduced scattering coefficient $\mu_{s}'$, from measurements of total reflectance and total transmittance using an integrating sphere setup~\citep{prahl1993determining}. 
The description of our phantom assembly as a three-layered samples with know pairs of thickness and refractive index for each layer was passed as input to the IAD program\footnote{\url{https://github.com/scottprahl/iad}}, along with a set of fixed parameters mostly describing the integrating sphere setup. The top and bottom layers described the glass coverslips, with a thickness of 0.15 mm and refractive index of 1.5. The middle layer described the tissue-mimicking phantom, with a thickness of 1.5 mm, a refractive index of 1.35 and scattering anisotropy of 0.85. 
The 2-stage IAD workflow was designed to better estimate optical properties in wavelength ranges where crosstalk is observed between measured absorption and reduced scattering coefficients. First, the IAD program was used to extract pairs of $\mu_{a}$ and $\mu_{s}'$ from pairs of total reflectance and total transmittance at each measurement wavelength. Second, the reduced scattering coefficient $\mu_{s}'$ spectra were fitted using a mixed Rayleigh-Mie model:
\begin{equation}
    \mu_{s}'=a'\left(f_{Ray}\left(\frac{\lambda}{\lambda_0}\right)^{-4}+\left(1-f_{Ray}\right)\left(\frac{\lambda}{\lambda_0}\right)^{-b_{Mie}}\right)
\label{eq:RayleighMie}
\end{equation}
where $a'=\mu_{s}'(\lambda_0)$, with $\lambda_0$ a reference wavelength fixed to 500 nm, $f_{Ray}$ is the fraction of Rayleigh scattering, $(1-f_{Ray})$ is the fraction of Mie scattering and $b_{Mie}$ is the Mie scattering power. Data points with wavelengths outside of the 450–1000 nm range were excluded. 

In biomedical optics, Rayleigh scattering refers to scattering by small scale particles much smaller than the wavelength (thus well-suited for shorter wavelengths), whilst Mie scattering is better suited for longer wavelengths given the microstructures found in biological tissues of interest. A combined Rayleigh-Mie scattering model is typically employed to account for contributions from both small-scale and wavelength-comparable scatterers in biological tissue, as neither Rayleigh nor Mie theory alone can adequately describe the tissue scattering behaviour across a broad range of measured wavelength \cite{Jacques2013OpticalReview,Shapey2022OpticalWindow}. In this study, we used mixed Rayleigh-Mie scattering model to reflect the heterogeneous particle sizes in tissue-mimicking phantoms, ranging from 50 - 600 nm for lipid particles to micron-scale haemoglobin clusters \cite{vanStaveren1991LightNm}.
Third, the IAD program was used a second time with $\mu_{s}'$ constrained using fitted values, thus obtaining revised values for $\mu_{a}$ and $\mu_{s}'$. 
For this study, the Windows version 3.15.0 of the IAD program was used.

The experimental setup described above measured total reflectance, corresponding to the sum of diffuse reflectance and specular reflectance. Models for fluorescence quantification generally use diffuse reflectance as input~\citep{Kim2010QuantificationMeasurements,Valdes2012ASurgery}. Specular reflectance was thus estimated using the Monte Carlo simulations, based on the known measurement geometry. The tool used for simulations is the Monte Carlo Multi-Layered (MCML) program. Input parameters matched those used for using the IAD program. A constant value of 0.04 was extracted by the MCML and subsequently subtracted to the total reflectance spectra to yield diffuse reflectance spectra.

Since the sampling of optical properties was coarser than that of the fluorescence emission, resampling was needed to proceed with processing steps. Linear interpolation was used to estimate optical properties over a 450–800 nm range with a 1 nm step, thus matching fluorescence measurements.

\subsection{PpIX concentration extraction}

\subsubsection*{Spectral unmixing}
A set of $n$ measured fluorescence emission spectra sampled over $m$ spectral bands can be represented in metric form as $V \in \mathbb{R}_+^{n \times m}$.
Each spectrum can be modelled as a linear combination of $r$ underlying spectral components, or endmembers. 
This representation defines a basis matrix $H \in \mathbb{R}_+^{r \times m}$, which contains the spectral components, and an abundance matrix $W \in \mathbb{R}_+^{n \times r}$, which encodes the contribution of each of the $r$ components in every of the $n$ measured emission spectra.
$V$ can thus be expressed as the product of $W$ and $H$:
\begin{equation}
    V \approx WH
    \label{eq:LinearDecomposition}
\end{equation}
As implied above, $W$, and $H$ are constrained to be non-negative, reflecting the physical nature of spectral intensities and components abundances.
Our primary objective is to quantify the abundance of PpIX in the presence of overlapping background components within our phantoms.
To this end, we seek accurate estimates of both the spectral components $H$ and the abundance matrix $W$, without prior knowledge of either.
This constitutes a blind spectral unmixing problem,
well suited to clinical settings where a priori information on tissue composition and in particular their fluorescence emission are not well defined: 
\begin{equation}
    \min_{W, H} \, D(V, WH)
    \label{eq:BlindUnmixing}
\end{equation}
where $D$ represents a distance measure that quantifies the dissimilarity between the measured data and the model.
The choice of distance measure is critical, as it should align with characteristics of the data and the desired result.
We chose to evaluate three distance measures:
1) Euclidean (L2) distance;
2) Kullback–Leibler (KL) divergence;
and 3) Hellinger distance.
Each offer different assumptions and sensitivities that influence the unmixing result.

The L2 distance 
is often a default choice and is indeed applied in similar works, however its usage implies the assumption that the measured data is subject to additive Gaussian noise. 
Specifically, the L2 distance arises from the negative log-likelihood under a Gaussian noise model
\begin{align}
    D_{\text{L2}}(V, WH) &= \|V - WH\|^2_2,
    \label{eq:L2Distance} \\
    P(V_{ij}|WH_{ij}) &= \frac{1}{\sqrt{2\pi\sigma^2}}\exp{-\frac{(V_{ij} - (WH)_{ij})^2}{2\sigma^2}},
    \label{eq:GaussianNoiseLikelihood} \\
    \mathcal{L}_{Gaussian} &= \sum_{ij}\frac{1}{2\sigma^2}(V_{ij} - (WH)_{ij})^2.
    \label{eq:GaussianNoise}
\end{align}
This model assume a uniform level of noise for all measurement, i.e. the model is homoscedastic. 

Being count based, our measured emission data is better described with a Poisson noise model (which is heteroscedastic by construction):
\begin{align}
    P(V_{ij}|(WH)_{ij}) &= \frac{(WH)_{ij}^{V_{ij}}e^{-(WH)_{ij}}}{V_{ij}!},
    \label{eq:PoissonNoiseLikelihood} \\
    \mathcal{L}_{Poisson} &= \sum_{ij}((WH)_{ij} - V_{ij}log((WH)_{ij}).
    \label{eq:PoissonNoise}
\end{align}
Instead of optimising the log-likelihood \eqref{eq:PoissonNoise} directly, non-negative matrix factorisation typically use both $V_{ij}$ and $(WH)_{ij}$ as parameters of Poisson distributions and use the Kullback-Leiber (KL) divergence to measure the difference between the two resulting distribution~\citep{nmf}.
The KL divergence between two Poisson distributions admits a well-known closed-form expression. Summing over all counts provide the following distance between $V$ and $WH$:
\begin{equation}
    D_{\text{KL}}(V| WH) = \sum_{ij} \left(V_{ij} \log\frac{V_{ij}}{(WH)_{ij}} - V_{ij} + (WH)_{ij} \right)
    \label{eq:KLDivergence}
\end{equation}

While Poisson noise is well suited for count based data, our measured spectra showed relatively high counts. 
Under this regime, the Poisson model is well approximated by a Gaussian, albeit with a variance proportional to the mean (now an heteroscedastic model).
This realisation allows us to exploit
the simpler and more efficient L2 objective after a data transformation aiming to reduce the impact of heteroscedasticity.
In this work, we took the square root of our data before running our unmixing with the L2 objective.
This is analogous to optimizing the Hellinger distance and we refer the resulting distance as $D_H$:
\begin{equation}
    D_{\text{H}}(V, WH) = \|\sqrt{V} - \sqrt{WH}\|^2_2
    \label{eq:HellingerDistance}
\end{equation}
where the square root is taken element-wise.

Due to the non-negativity constraint, we adopted a non-negative matrix factorisation (NMF) algorithm to perform the optimisation.
While our problem is non-convex in $W$ and $H$, it is convex for each variable individually.
As such, this optimisation is typically performed using interleaved updates for $W$ and $H$ which are constructed to preserve non-negativity.
Multiplicative update rules were previously derived in~\citep{nmf} both for the Euclidean distance measure
\begin{equation}
    H_{ij} \leftarrow H_{ij} \frac{(W^\top V)_{ij}}{(W^\top W H)_{ij}}, \quad\quad
    W_{ij} \leftarrow W_{ij} \frac{(V H^\top)_{ij}}{(W H H^\top)_{ij}}
    \label{eq:UpdateRuleEuclidian}
\end{equation}
and for the KL divergence measure 
\begin{equation}
    H_{ij} \leftarrow H_{ij} \frac{\sum_k{W_{ki}V_{kj}/(WH)_{kj}}}{\sum_lW_{la}}, \quad\quad
    W_{ij} \leftarrow W_{ij} \frac{\sum_k{H_{ki}V_{kj}/(WH)_{kj}}}{\sum_lH_{la}}
    \label{eq:UpdateRuleKL}
\end{equation}
For the Hellinger case, we simply use the Euclidean update rule \eqref{eq:UpdateRuleEuclidian} on the data transformed by the square root.
These updates were formulated to guarantee convergence to a local minimum whilst maintaining non-negativity and were significantly more efficient than standard gradient descent.
While we were not directly interested in the residual errors of our unmixing models, it is helpful when comparing methods to choose a single consistent metric irrespective of the underpinning model fitting assumptions.
In this work, we chose to use the normalised L2 distance for reporting and comparison:
\begin{equation}
    D_{\text{L2'}}(V, WH) = \frac{\|V - WH\|^2_2}{\|V\|^2_2}.
    \label{eq:ResidualsL2Distance}
\end{equation}

\subsubsection*{Regularisation}
The NMF models above assume independence across all individual measurements.
In practice, it can however be expected that the 
variables in $H$ are in fact highly correlated with their spectral neighbours.
In other words, the spectral signatures in $H$ should generally be smooth functions of $\lambda$. 
We modified the cost function \eqref{eq:BlindUnmixing} to reflect this by incorporating a Tikhonov regularisation term based on the finite difference along the spectral dimension: 
\begin{equation}
    \min_{W,H}\bigg[D(V,WH) + \alpha_\Gamma\|\Gamma H\|_2^2\bigg] \quad \text{where} \quad
    \Gamma = 
    \begin{bmatrix}
    1 & -1 & 0 & \cdots & 0 \\
    0 & 1 & -1 & \cdots & 0 \\
    \vdots & \ddots & \ddots & \ddots & \vdots \\
    0 & \cdots & 0 & 1 & -1
    \end{bmatrix}
    \label{eq:Regularisation}
\end{equation}
where $\alpha_\Gamma$ is a scalar that controls the amount of regularisation. In order to avoid derivation of a novel multiplicative update rule, we opted to optimise for our regularisation term using alternating optimisation. This allowed us to utilise the highly stable and efficient multiplicative update rule whilst additionally optimising for our regularisation term.
Specifically we applied a standard gradient descent step after applying the multiplicative update rule. The output was clipped to ensure non-negativity. Our update rule for $H$ then becomes: 
\begin{equation}
    H_{ij} \leftarrow \max \bigg(0, H_{ij}M - \alpha_\Gamma (\Gamma^T \Gamma H)_{ij} \bigg)
    \label{eq:UpdateRuleRegularised}
\end{equation}
where $M$ represents the multiplicative factor for $H$ from the update rules above. With $\alpha_\Gamma$ chosen to be small enough, we found this to be an effective approach that compromised exactness for improved convergence. This method would also allow the extension to any arbitrary, but differentiable, regularisation terms without the need to derive a strict update rule. 

Additionally, in our dataset, we systematically observed a sharp feature with variable intensity located at 690 nm, which we could not attribute to meaningful fluorescence emission from any constituent in the phantoms. Rather than attempting to remove this using pre-processing, we instead modelled it in our blind unmixing. 
To do this, we initialised an additional endmember as a sharp peak around 690 nm. We did not apply any smoothness regularisation to this endmember during optimisation. This approach could effectively model out systematic noise or other high frequency features which otherwise corrupt the rest of the components in our unmixing.

\subsubsection*{PpIX quantification}
While our unmixing provides relative abundances of PpIX, these need to be calibrated against the known concentrations in our phantom dataset to provide meaningful predictions. However, the available ground-truth concentration corresponds to the total PpIX concentration, as defined by the phantoms design. The extracted abundances of $\text{PpIX}_{635}$ and $\text{PpIX}_{620}$ therefore need to be recombined to estimate a total PpIX concentration from measured emission. 
First we selected all endmembers in $H$ that correspond to PpIX fluorescence.
This was done by thresholding the KL-divergence between the endmember and a simplified bi-Gaussian model of PpIX fluorescence defined a priori from literature values.
This provided $q$ end members spectra represented as a matrix $H^{ppix} \in \mathbb{R}^{q\times m}$ and corresponding abundances $W^{ppix} \in \mathbb{R}^{n\times q}$.
We formed our prediction through the linear combination of these abundances weighted by a weight vector $\kappa\in\mathbb{R}^q$.
Calibration was then performed using linear least squares to optimise for the weights: 
\begin{equation}
    \min_\kappa\|y-W^{ppix}\kappa \|_2^2
    \label{eq:PpIXCalibration}
\end{equation}

\subsection{Optical properties correction}
As detailed in Section~\ref{sec:optcorrbkg},
various approaches to correct for optical distortions have been proposed, whether empirical or model-based.
Here, we adapt and expand a validated empirical approach~\citep{Valdes2012ASurgery} consisting of a dual-band normalization of the raw fluorescence emission:
\begin{equation}
V'(\lambda) = \frac{V(\lambda)}{R_{ex} R_{em}^\alpha}
\label{eq:ValdesCorrectionOriginal}
\end{equation}
where $R_{ex}$ and $R_{em}$ are found by integrating the reflectance across defined ranges for the excitation and emission light. 
The value of $\alpha$ is empirically derived (or calibrated) to optimise linearity against known PpIX concentrations.
In \citet{Valdes2012ASurgery}, $R_{em}$ is extracted by integration over 625-645 nm corresponding to the primary emission peak of PpIX, and $R_{ex}$ by integration over 465-485 nm.
We adopt this method as our baseline with a minor change to shift the integration range for $R_{ex}$ to be 430-450nm as we have robust data available in this range, closer to the true excitation wavelength wavelength of 405 nm, similarly to previously published work \citep{Xie2017Wide-fieldResection}.

We noted that, as this correction reduced down to a scaling factor, it was mathematically equivalent to apply this scaling either pre-unmixing to the measured raw spectra, or post unmixing to the extracted abundances, provided that the unmixing process is linear.
In light of this observation, we explored two competing ways how we might expand this optical correction method which we termed \emph{spectral correction} and \emph{abundance correction}.

Additionally, our spectral correction method applied a wavelength-dependent scaling factor instead of a constant scaling factor:
\begin{equation}
V'(\lambda) = \frac{V(\lambda)}{R_{ex} R(\lambda)^\alpha}
\label{eq:SpectralCorrection}
\end{equation}
These corrected spectra were then used as input to the unmixing as detailed before.
Conversely, our abundance correction method leveraged the availability of exact component spectra post-unmixing to calculate a correction factor specific to each fluorophore, in particular each form of PpIX. The corrected abundances $W'_{i}$ were obtained by adapting the dual-band normalization method, as expressed in \eqref{eq:AbundanceCorrection}. $\bar{R}_{em,i}$ was thus calculated as weighted average reflectance for the corresponding endmember spectrum $H_i(\lambda)$, matching the spectral shape of fluorescence emission and corresponding diffuse reflectance spectrum:
\begin{equation}
W_{i}' = \frac{W_{i}}{R_{ex} \bar{R}_{em,i}^\alpha} \quad \text{where} \quad \bar{R}_{em,i} = \frac{\int R(\lambda) \cdot H_{i}(\lambda) \, d\lambda}{\int H_{i}(\lambda) \, d\lambda}
\label{eq:AbundanceCorrection}
\end{equation}

\begin{figure}[!htb]
  \centering
  \includegraphics[width=0.8\linewidth]{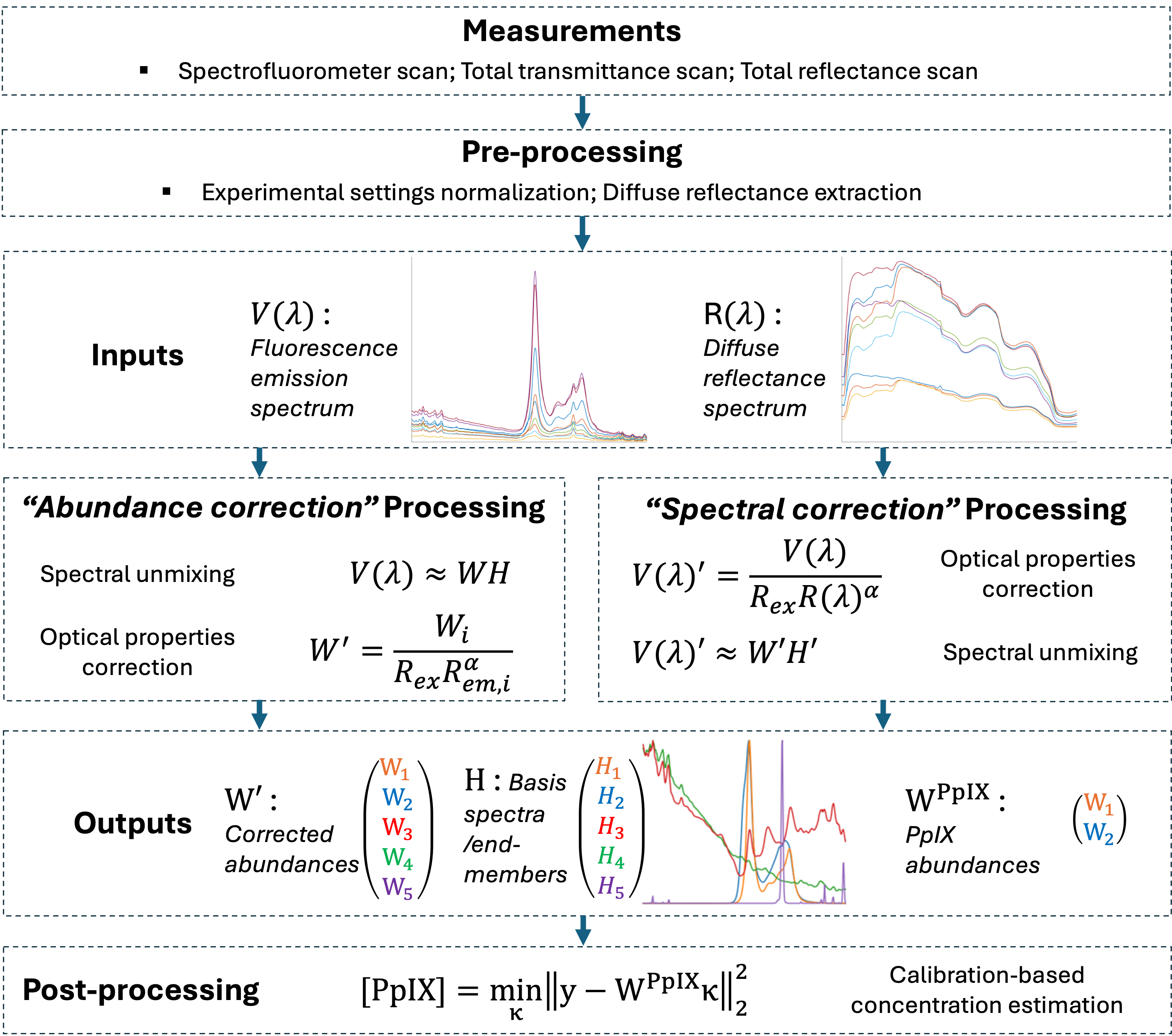}
  \caption{An illustration of the overall processing workflow including spectral unmixing and concentration estimation.}
  \label{fig:Conceptual_workflow_figure}
\end{figure}

In our work, the calibration of $\alpha$ was performed similarly to the original method~\citep{Valdes2012ASurgery}. The data were processed with values of $\alpha$ ranging from -2 to 2. After quantification, the optimal value for $\alpha$ was determined by assessing goodness-of-fit versus ground-truth PpIX concentration. 
An overall processing workflow is shown in \figref{fig:Conceptual_workflow_figure}.

\section{Results}

\subsection{Phantom properties}
A panel of 108 phantoms was used in this study. These correspond to 3 replicates of each combination of IL concentrations of 1, 3 and 5\%, Hb concentrations of 0.5, 1.25 and 2 mg/mL and PpIX concentrations of 0, 5, 10 and 15 µg/mL. 

\begin{figure}
    \centering
    \includegraphics[width=0.8\linewidth]{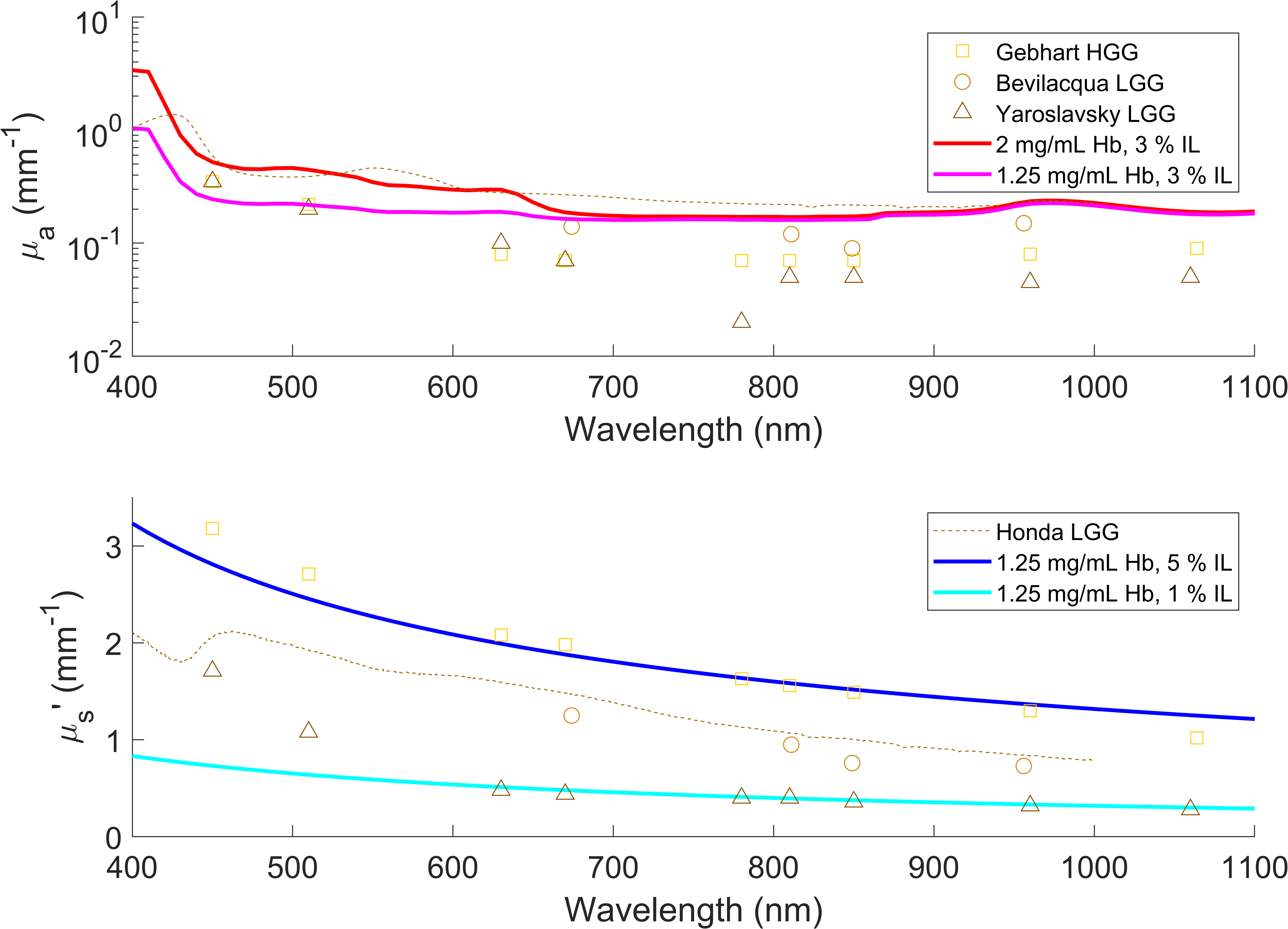}
  \caption{Comparison of optical properties (\textit{Top}: Absorption coefficient $\mu_a$, \textit{Bottom}: Reduced scattering coefficient $\mu'_{s}$) between measurements of a selection of our phantoms as plotted in solid lines and reported values for glioma tissues from various sources as presented in respective discrete symbols, over a wavelength range of 400 - 1100 nm \citep{Yaroslavsky2002OpticalRange,Gebhart2006InAdding-doubling, DeRicBevilacqua1999InBrain, Honda2018DeterminationAcid}.}
    \label{fig:phantoms_OPs}
\end{figure}

In turn, absorption coefficients and reduced scattering coefficients covering ranges of 0.2 – 0.6 mm$^{-1}$ and 0.5 – 3 mm$^{-1}$ were respectively measured at 500 nm. As pictured in \figref{fig:phantoms_OPs}, these matched reported ranges of optical properties encountered in brain and tumour tissue \citep{Yaroslavsky2002OpticalRange,Gebhart2006InAdding-doubling,DeRicBevilacqua1999InBrain, Honda2018DeterminationAcid}. Most of these data points were found to be compiled and made publicly available \citep{Shapey2022OpticalWindow}.
The use of Hb as an optical absorbing agent in these phantoms allowed for dramatic variations of $\mu_{a}$ with wavelength, similar to typical human brain and tumour tissue. At 400 nm, Hb features significantly stronger absorption, resulting in absorption coefficients of 1-3.5 mm$^{-1}$. 
Inversely, at 630 nm which is close to the maximum emission wavelength of PpIX, absorption coefficients of 0.15-0.35 mm$^{-1}$ were measured.
The use of IL as an optical scattering agent achieved reduced scattering coefficients mainly following a Mie power law, with measured average ($\pm$standard deviation) fraction of Mie scattering $(1-f_{Ray})$ = 0.066 ($\pm$0.042) and Mie scattering power $b_{Mie}$ = 0.88 ($\pm$0.15). 

\begin{figure}
    \centering
    \includegraphics[width=0.9\linewidth]{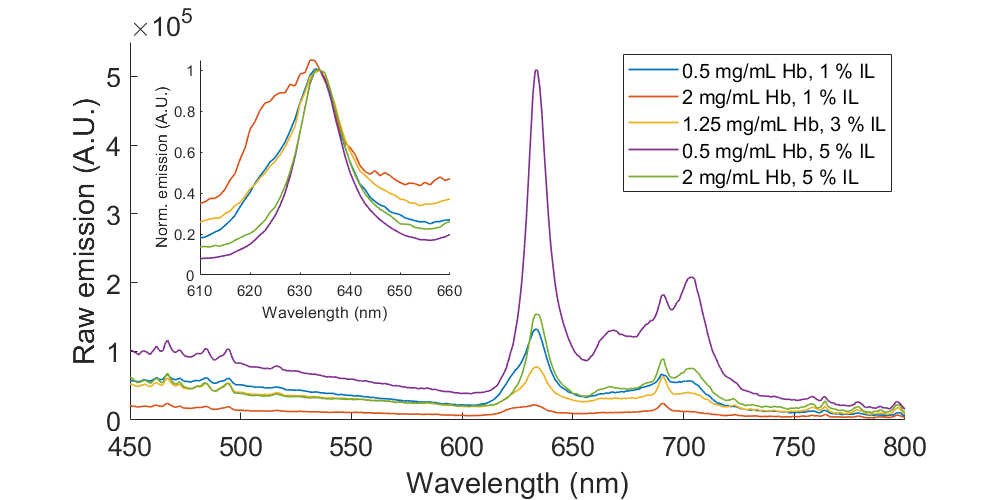}
    \caption{Raw fluorescence emission spectra for a selection of our phantoms spanning various concentrations of optical agents IL and Hb, with fixed PpIX concentration of 5 $\mu$g/mL. The overall intensity and spectral shape of PpIX fluorescence emission is markedly impacted by the concentrations of optical agents. An insert shows normalized intensities at 634 nm, focusing on the primary emission peak of PpIX. The orange, blue and yellow curves show significant presence of PpIX$_{620}$.}
    \label{fig:phantoms_fluorescence}
\end{figure}

Stock PpIX solution diluted in 1x PBS showed typical fluorescence emission from PpIX$_{635}$ under 405 nm excitation. Linear increase in intensity of fluorescence emission at 634 nm was observed for dilutions with PpIX concentration over a range of 0-100 µg/mL ($R^2$ = 0.999). 
In phantoms however, PpIX fluorescence emission was found to vary not only in intensity but also in spectral shape with phantom composition, depending on IL, Hb and PpIX concentrations. The resulting fluorescence emission can be described as mixture of emission from both PpIX$_{620}$ and PpIX$_{635}$, plus background emission. This matched behaviour previously described in the literature, both in phantoms and human tissue~\citep{Alston2018NonlinearPhantoms,SueroMolina2023UnravelingBiology,Korner2024}.  \figref{fig:phantoms_fluorescence} illustrates this phenomenon by showing raw fluorescence emission spectra for a selection of our phantoms with varying concentrations of Hb and IL and fixed PpIX concentration of 5 $\mu g/mL$. An insert shows normalized intensities at 634 nm, restricted to a wavelength range of 610 - 660 nm, encompassing the primary emission peak of PpIX. Significant presence of PpIX$_{620}$ is thus noticeable in the orange, blue and yellow curves. Background fluorescence emission was observed in all phantoms and was attributed to both gelatin and IL. A sharp 690 nm peak also observed in all phantoms with varying intensity was attributed to IL. 

\subsection{PpIX quantification}
\begin{figure}[htb]
    \centering
    \includegraphics[width=1.0\linewidth]{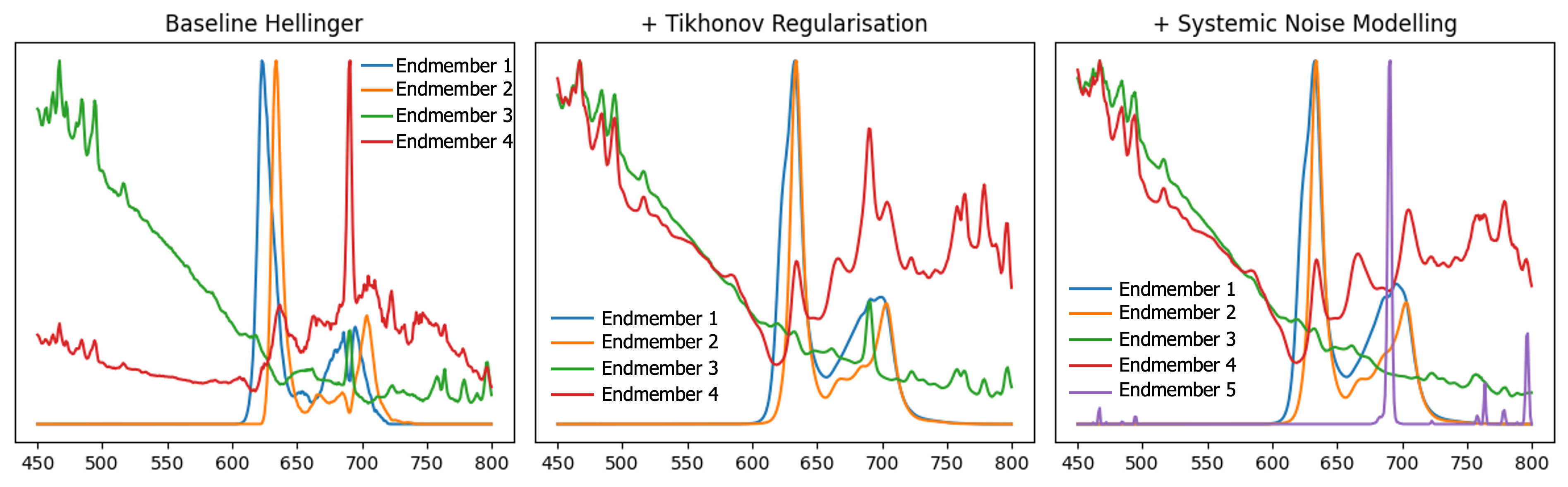}

\caption{Visualisation of the basis spectra - endmembers discovered using the proposed unmixing approach with the Hellinger cost function and different regularisation strategies. Left: Spectral endmembers identified without regularisation. Middle: Tikhonov regularisation is applied to promote smoother endmembers. Right: A fifth endmember is introduced to capture a sharp systematic feature with the additional high-frequency systemic noise modelling. All endmembers are normalised for visualisation purposes.}
    \label{fig:regularisation}
\end{figure}

\figref{fig:regularisation} shows the evolution of our basis spectra discovered as our unmixing regularisations were added. The smoothness regularisation allowed our basis to represent more plausible spectral signatures by ignoring high frequency noise, and sharp features modelling can be seen to absorb the sharp peak from the green and red endmembers.

\begin{figure}[htb]
    \centering
    \includegraphics[width=0.85\linewidth]{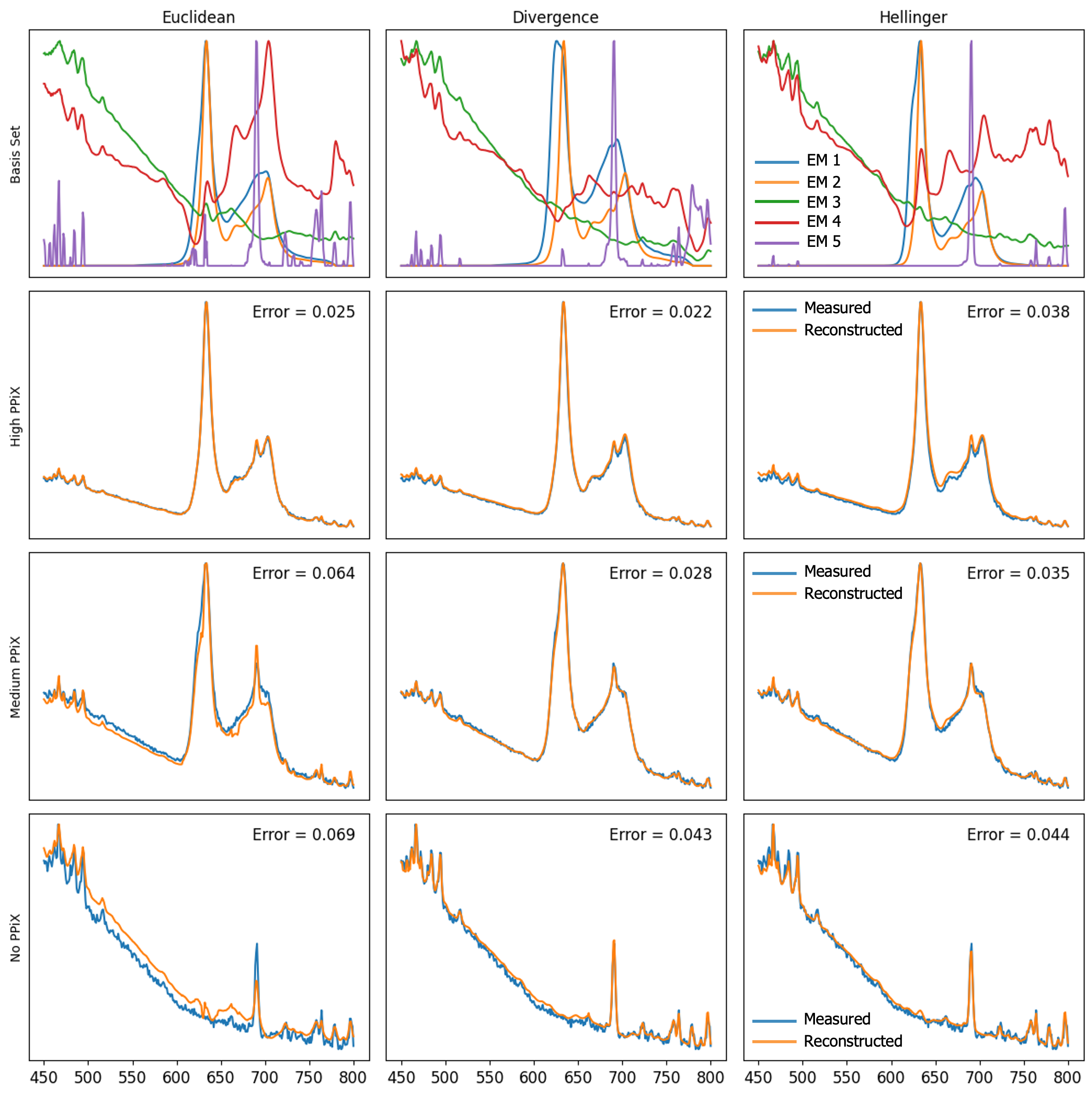}
   
   \caption{Visualisation of the discovered spectral endmembers and some exemplar reconstructions for each of our unmixing methods along side the measured spectra. Each endmember (EM) is normalised and the Hellinger basis is squared, for visualisation purposes. The reconstruction errors are also displayed for quantitative comparison. Reconstruction error (Euclidean distance) is provided for each sample.}
    \label{fig:distance-measure-results}
\end{figure}

\figref{fig:distance-measure-results} shows a comparison of our regularised bases for each distance measure, along with some examples of reconstructions. The $L_2$ reconstruction error of different unmixing methods and regularisation approaches is shown in \tabref{tab:nl2-scores}.  

\begin{table}[htb]
\centering

\caption{Normalised $L_2$ reconstruction errors for our different unmixing objectives and regularisation improvements. We show the mean and standard deviation over five runs to account for variability in the unmixing process.
\label{tab:nl2-scores}}
\small
\begin{tabular}{l|ccc}
\hline
\begin{tabular}[c]{@{}l@{}}Unmixing\\Objective\end{tabular} 
& Baseline     & + Tikhonov regularisation   &
\begin{tabular}[c]{@{}l@{}} + Tikhonov regularisation \\ + System noise modelling \end{tabular}  \\
\hline
Euclidean   & 0.0169±.0005   & 0.0258±.0001     & 0.0243±.0001   \\
Divergence  & 0.0207±.0002   &  0.0318±.0001   &   0.0310±.0002  \\
Hellinger   &  0.0179±.0002  &  0.0271±.0001  &  0.0263±.0000 \\
\hline
\end{tabular}
\end{table}

\begin{figure}[!htb]
  \centering
  \includegraphics[width=0.5\linewidth]{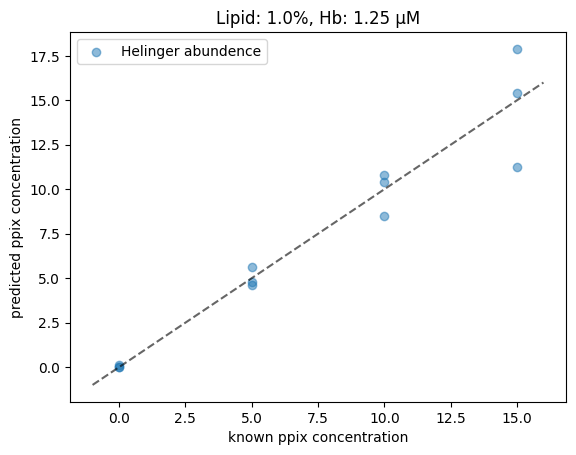}
 
 \caption{Representative PpIX reconstruction of one phantom composition.}
  \label{fig:example_correlation}
\end{figure}

We found the Hellinger distance measure, along with our abundance correction, provided the best correlation of predicted and ground truth PpIX concentration in our phantom dataset, achieving $R^2$ of $0.918\pm.002$. A representative phantom example to demonstrate the correlation between the true concentrations and the predicted ones by Helinger abundance approach is shown in \figref{fig:example_correlation}. Using this method we found $\alpha=-1.0$ to be optimal. 
It is of note that performance rankings remained consistent for any given correction or unmixing method.

As shown in \tabref{tab:nl2-scores}, the performance of fluorescence emission spectrum reconstruction using different unmixing endmember approaches is broadly comparable, with the $L_2$ reconstruction errors ranging from 0.0169 for the Baseline-Euclidean approach to 0.0318 for the Divergence-based regularised method. However, spectral reconstruction accuracy is not the primary objective of this study; instead, the main aim is the accurate estimation of PpIX concentration. From this perspective, the Hellinger unmixing method shows the strongest agreement between predicted and known PpIX concentrations across all phantom conditions, as reflected by the highest coefficient of determination ($R^2$ = 0.928) and the lowest root mean square error (RMSE = 1.652), as shown in the correlation performance results in \tabref{tab:correlation-scores} and   \tabref{tab:rmse-scores}).
This improved performance may stem from the inherent robustness properties of the Hellinger metric, which moderates the influence of large-intensity deviations and reduces sensitivity to heteroscedastic noise, thereby enabling more stable and reliable estimation of low-abundance components such as PpIX.

\begin{figure}[htb]
    \centering
    \includegraphics[width=0.8\linewidth]{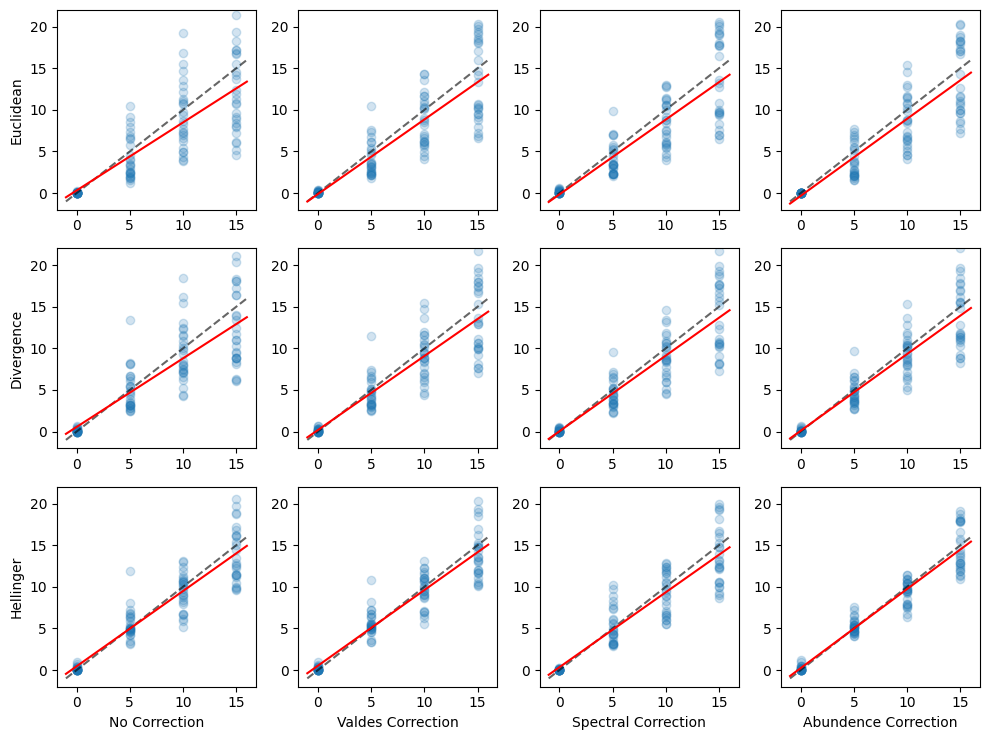}
     
   \caption{The scatter plots demonstrate correlations between the predicted (x-axis) and known PpIX concentrations (y-axis) in our entire phantom dataset for each of our unmixing and optical property correction methods. The dashed reference line represents the ideal 1:1 relationship, while the red line indicates the line of best fit through the predictions. In the scatter plot, regions of darker color indicate overlapping data points, corresponding to multiple samples with identical predicted values.}
    \label{fig:correlation-visualisation}
\end{figure}

 A visualisation of the correlations between the predicted and known PpIX concentrations in our entire phantom dataset for each of our unmixing and optical property correction methods is shown in \figref{fig:correlation-visualisation}. The dashed reference line represents the ideal 1:1 relationship, while the red line indicates the line of best fit through the predictions. The Hellinger distance together with abundance correction demonstrates the closest 1:1 correlation.
\begin{table}[htb]
\centering
\caption{$R^2$ values for our PpIX concentrations predictions for different unmixing cost functions (Euclidian, Divergence, Hellinger) and optical property correction techniques (Baseline, Regularised, Valdes, Spectral, Abundance). Mean and standard deviation over five runs are shown to account for variability in the unmixing process.\label{tab:correlation-scores}}
\small
\begin{tabular}{l|ccccc}
\hline
\begin{tabular}[c]{@{}l@{}}Unmixing\\Objective\end{tabular} 
& \begin{tabular}[c]{@{}l@{}}No correction\\Baseline \end{tabular}  &\begin{tabular}[c]{@{}l@{}} No correction \\Regularised \end{tabular} &\begin{tabular}[c]{@{}l@{}} Valdes\\ correction  \end{tabular}    &\begin{tabular}[c]{@{}l@{}} Spectral\\ correction \end{tabular}   &\begin{tabular}[c]{@{}l@{}} Abundance\\ correction \end{tabular}  \\
\hline
Euclidean   & 0.497±.054   & 0.645±.036   & 0.737±.004   & 0.754±.008   & 0.771±.036   \\
Divergence  & 0.586±.015   & 0.689±.001   & 0.766±.010   & 0.790±.002   & 0.826±.001   \\
Hellinger   & 0.809±.018   & 0.846±.004   & 0.865±.038   & 0.902±.012   & \equalbold{0.918±.002} \\
\hline
\end{tabular}
\end{table}

\begin{table}[htb]
\centering

\caption{RMSE values (µg/mL) for our PpIX concentrations predictions for different unmixing cost functions (Euclidean, Divergence, Hellinger) and optical property correction techniques (Baseline, Regularised, Valdes, Spectral, Abundance). Mean and standard deviation over five runs are shown to account for variability in the unmixing process.\label{tab:rmse-scores}}
\small
\begin{tabular}{l|ccccc}
\hline
\begin{tabular}[c]{@{}l@{}}Unmixing\\Objective\end{tabular} 
& \begin{tabular}[c]{@{}l@{}}No correction\\Baseline \end{tabular}  &\begin{tabular}[c]{@{}l@{}} No correction \\Regularised \end{tabular} &\begin{tabular}[c]{@{}l@{}} Valdes\\ correction  \end{tabular}    &\begin{tabular}[c]{@{}l@{}} Spectral\\ correction \end{tabular}   &\begin{tabular}[c]{@{}l@{}} Abundance\\ correction \end{tabular}  \\
\hline
Euclidean   & 5.290±.576 & 3.793±.151 & 3.376±.033 & 3.155±.052 & 3.120±.338 \\
Divergence  & 4.136±.143 & 3.350±.061 & 3.292±.099 & 2.714±.023 & 2.516±.160 \\
Hellinger   & 2.784±.239 & 2.243±.023 & 2.207±.023 & 2.073±.240 & \textbf{1.652±.011} \\
\hline
\end{tabular}
\end{table}

\section{Discussion}

Robust quantification of PpIX fluorescence in glioma has the potential to measurably improve treatment outcomes for tumour patients, yet remains a significant medical imaging challenge. The development of validated quantification workflows has been constrained by the lack of reproducible phantoms that simultaneously replicate two major challenges to objective PpIX detection, namely optical distortion and PpIX micro-environment interactions \citep{Walke2023ChallengesMeasurements}. This is compounded by the interdependent relationship between these two, especially in vivo, with LGG showing distinct PpIX emission and optical properties characteristics versus HGG \citep{Gebhart2006InAdding-doubling,Alston2019SpectralGliomas}.

Our proposed work utilised novel, tissue mimicking, dual emission state PpIX phantoms to develop and validate an unmixing pipeline that successfully identifies two prominent forms of PpIX fluorescence in complex environments with overlapping and clinically relevant fluorophores.
Using our optical correction method we are then able to show a strong correlation of a unified predicted PpIX concentration in a pipeline capable of adapting to varying and unanticipated micro-environments. 

The use of blind unmixing techniques enables this method to account for unknown contributions of background autofluorescence, and is not dependent on a priori knowledge of overlapping fluorophores known to show significant inter and intratumoural heterogeneity in glioma \citep{Badr2020MetabolicTumors}. Additionally, our proposed spectral unmixing method can be adapted to process data from both point-based and widefield devices, despite their varying sampling rates and spectral resolutions. Such adaptability is critical for translation to in vivo imaging, where trade off between resolution and workflow integration are significant barriers to the adoption of quantitative imaging technologies \citep{Kotwal2024HyperspectralApplications}.

The validation of this technique is enabled by our novel tissue mimicking phantoms, which display variable, dual-state PpIX fluorescence emission with optical properties matched to glioma tissue. The phantom composition included IL and Hb to mimic biological tissue composition and associated optical properties over the wavelength range of interest, encompassing wavelength ranges of excitation and emission for PpIX. While we do not control the mix of oxy- and deoxy-haemoglobin within the tissue, the optical correction model is calibrated using the measured phantom optical properties and therefore is self-adjusted for the phantom composition. Although the individual components of the phantoms are widely used in the field, this exact composition, moreover in a solid form is, to the best of our knowledge, novel. Additionally, we describe a novel larger framework for preparing samples and measuring these using two separate devices, producing paired fluorescence and optical properties measurements. 
While our study focused on mimicking glioma tissue, the phantom composition can be adjusted to mimic other types of tissue, for example, white or gray matter, or other tumour types. Using a richer composition with various absorbers would allow reproduction of an even larger range of tissue types. The gelatin base allows for easy handlings, and potentially enables shaping to create inclusions or anatomical features in more complex phantoms. Indeed, while our study relied on a simple homogenous, mono-layered fluorescent tissue-mimicking phantom model, expanding the phantom design would allow for mimicking more realistic settings, including multi-layered structures or depth-resolved fluorophore inclusions. Going forward, our proposed experimental setup measuring paired fluorescence and optical properties could be used as presented for ex vivo preclinical or clinical studies.

Our phantoms exhibited dual-state PpIX fluorescence emissions and fluorescence background analogous to tissue autofluorescence (attributed to IL and gelatin) was observed in all phantoms. More complex fluorescence backgrounds could be introduced by introducing additional fluorophores to the phantom composition. 
 We acknowledge that PpIX fluorescence can be influenced by the pH of the surrounding medium. In this study, the pH of the gelatin-based solid phantoms could not be measured using standard pH probes. By using buffered PBS with standard pH of 7.4, we would expect the resulting solid phantoms with a physiologically relevant pH value between 7 - 8. Although direct pH measurements were not feasible, all phantoms were fabricated using the same batch of materials, and PpIX stock solution, following an identical established preparation protocol. Therefore, any pH-related effects are expected to be consistent across samples and unlikely to bias analyses across these phantoms. Future work will explore alternative approaches to directly control and assess pH distributions in solid optical phantoms.  Futhermore, previous studies have shown that lipid association and partial aggregation of PpIX lead to a blue-shifted emission band around ~620 nm, whereas monomeric PpIX in aqueous environments retains the characteristic ~635 nm emission \citep{Alston2018NonlinearPhantoms,SueroMolina2023UnravelingBiology}. In the work of \citep{Alston2018NonlinearPhantoms}, when the volume fraction (\%) of Intralipid in the phantom is between 0.005 - 0.015, the two PpIX emission states can be observed.

A sharp peak located at 690 nm visible in most spectra was investigated. It was found to shift with excitation wavelength and was attributed to IL alone. This was  considered as a separate contribution in our fluorescence data and modelled accordingly. The versatility of our proposed unmixing method enabled the separation of all these contributions to the detected fluorescence emission, thereby yielding accurate PpIX quantification.

 The proposed quantification methods were based on the optical correction of fluorescence signal and spectral unmixing for extraction of PpIX contribution. We adapted an empirical correction model using diffuse reflectance and fluorescence emission as inputs, which was calibrated using the presented panel of tissue-mimicking phantoms. This calibration - being derived from intrinsic properties of the phantoms such as reflectance -is therefore directly linked to the molecular composition and microstructure within the samples. Thus, the model is limited to similar samples, and should therefore be recalibrated for application to samples with a different composition. 
Similarly to most published optical correction methods, the integration range for estimating $R_{ex}$ did not match the known 405 nm excitation. In helping with Xie et al., we chose the 430-450 nm range as it was the most reliable wavelenght range closest to the actual excitation. Other methods have used ranges up to 465-485 nm \citep{Valdes2012ASurgery}, and alternative approaches for extrapolating the values of optical properties at 405 nm have also been proposed \citep{Kim2010QuantificationMeasurements}.
Even though our experimental workflow does yield quantitative optical properties extraction, we have not implemented model-based corrections. Indeed, these generally rely on the assumption of a semi-infinite depth for the samples \citep{Kim2010QuantificationMeasurements}, which is not compatible with our experimental setup measuring samples with fixed 1.5 mm thickness.

High correlation between predicted and known PpIX concentration was obtained ($R^{2}$=0.918), showing good performance of our proposed processing pipeline. Nevertheless, a significant residual error in PpIX concentration predictions was observed and could be accounted for by a combination of factors. 
Given manual manufacturing process, the ground-truth PpIX concentration may show some variance. Small local heterogeneities such as air bubbles might exist in the phantoms due to the mixing and pouring processes, impairing both fluorescence and optical properties readings. Additionally, the collection of total reflectance and total transmittance data is very sensitive to the quality of sample preparation and any misalignment in the integrating sphere setup, thus introducing potential errors. Finally, whilst spectral mixing is assumed to be linear, there could be potential non-linear interactions that would not be captured by our linear unmixing.

\section{Conclusion}

In this work, we developed and validated a novel dual-state PpIX quantification pipeline able to accurately reconstruct PpIX contribution across the range of optical and photochemical properties seen in low and high grade glioma. Our proposed workflow includes correction of optical properties-induced distortions using an adapted empirical model, and spectral unmixing of fluorescence emission. This blind unmixing method successfully detected two PpIX photostates and background emission components in tissue-mimicking phantoms. The described pipeline in transferable to tissue investigation, thus paving the way to preclinical and clinical studies. Additionally, our workflow uniquely utilises white-light diffuse reflectance and fluorescence emission as input, and can therefore be adapted to a wide-field imaging implementation. This novel integration lowers barriers to deployment and strengthens the case for future clinical translation.

\FloatBarrier

\section*{Acknowledgements}
JS and TV are co-founders and shareholders of Hypervision Surgical Ltd London, UK.
The authors have no other relevant interests to declare.

This project received funding from the Wellcome Trust under their Innovation Award program (Grant Reference Number WT223880/Z/21/Z).
This work was supported by core funding from the Wellcome/EPSRC [WT203148/Z/16/Z; NS/A000049/1].
The views expressed are those of the author(s) and not nescessarily those of the Wellcome Trust. 
This project received funding by the National Institute for Health and Care Research (NIHR) under its Invention for Innovation (i4i) Programme [NIHR202114].
The views expressed are those of the author(s) and not necessarily those of the NIHR or the Department of Health and Social Care.
For the purpose of open access, the authors have applied a CC BY public copyright license to any Author Accepted Manuscript version arising from this submission.

\printbibliography

\end{document}